\documentstyle[aps,12pt,preprint,amssymb]{revtex}
\tightenlines
\begin{document}
\title{{\bf Non-Chern-Simons Topological Mass Generation in (2+1) Dimensions}}
\author{Deusdedit M. Medeiros\thanks{%
Permanent address: Universidade Estadual do Cear\'{a}-UECE, CECITEC, Rua
Solon Medeiros, s/n, 63660-000 Tau\'{a}-Ce, Brazil}, R. R. Landim,\ C. A. S.
Almeida}
\address{{\normalsize {\it Universidade Federal do Cear\'{a}}}\thinspace \\
{\normalsize {\it Physics Department}}\\
{\normalsize {\it C.P. 6030, 60470-455 Fortaleza-Ce, Brazil}}\thanks{%
Electronic address: dedit@fisica.ufc.br, renan@fisica.ufc.br,
carlos@fisica.ufc.br}} \maketitle

\begin{abstract}
By dimensional reduction of a massive B$\wedge $F theory, a new
topological field theory is constructed in (2+1) dimensions. Two
different topological terms, one involving a scalar and a
Kalb-Ramond fields and another one equivalent to the
four-dimensional B$\wedge $F term, are present. We constructed two
actions with these topological terms and show that a topological
mass generation mechanism can be implemented. Using the
non-Chern-Simons topological term, an action is proposed leading
to a classical duality relation between Klein-Gordon and Maxwell
actions. We also have shown that an action in (2+1) dimensions
with the Kalb-Ramond field is related by Buscher's duality
transformation to a massive gauge-invariant St\"{u}ckelberg-type
theory.
\end{abstract}

\newpage
\section{\bf Introduction}
\vspace*{-0.5pt}

\noindent It is well known that a topological Chern-Simons term
give rise to gauge invariant mass to the gauge field
\cite{jackiw1}. In (3+1) dimensions, two procedures are generally
used for generating massive gauge fields consistent with
gauge-invariance. One is the St\"{u}ckelberg formulation
\cite{stuck} which is the more familiar Higgs mechanism in its
simplest form and the other one uses a 2-form potential (Kalb-Ramond field) $%
B$ coupled to the one-form gauge potential $A$ through a $B\wedge
F$ term, where $F=dA$ is the field-strength of $A$
\cite{kalb,cremer,allen}.

This theory has two types of gauge-invariance and has therefore
highly constrained couplings and is very geometrical. In a first
order formulation
of the non-Abelian Yang-Mills gauge theory ( BF-YM model) \cite{fucito} a $%
B\wedge F$ term has been used to contribute for a discussion of
quark confinement in continuum QCD. Additionally, transmutation of
statistics of point particles in (2+1) dimensions can be
generalized to that of strings in (3+1) dimensions via a $B\wedge
F$ term \cite{gambini}. On the other hand, the so-called mixed
Chern-Simons term, which involves two one-form gauge fields, was
recently studied in connection with certain condensed matter
systems \cite{diamantini}, namely {\it Josephson junction arrays}
\cite{eckern}.

In this letter, we consider a new topological term in (2+1)
dimensions, obtained by dimensional reduction from a $B\wedge F$
(3+1) dimensional Abelian model (to the best of our knowledge,
this term has not been studied in the explicit form presented
here). This term has the form $B\wedge d\phi$ and involves $B$ and
a 0-form field $\phi$. We show that this term can generate mass
for the Kalb-Ramond field as well as for the scalar field. An
action with this non-Chern-Simons topological term, leads us to a
classical duality between the free Klein-Gordon and Maxwell
actions. On the other hand, using the St\"{u}ckelberg formulation,
an alternative massive gauge-invariant model is constructed.
Finally, motivated by the fact that a interchange between
topological and Noether current usually denotes a duality
transformation, we also shown that a topological action in 2+1
dimensions with the Kalb-Ramond field is related by Buscher's
duality transformation \cite{buscher} to a St\"{u}kelberg-type
theory.

%\pagebreak

\textheight=7.8truein \setcounter{footnote}{0}
\renewcommand{\thefootnote}{\alph{footnote}}

\section{\bf Non-Chern-Simons Topological Term}

\noindent Our starting point is the Abelian $B\wedge F$
four-dimensional action \cite{guada,cattaneo}

\begin{equation}
S_{BF}=\int_{M_4}\left\{ B\wedge F-g^2B\wedge \,^{*}B\right\}.
\label{bfm01}
\end{equation}

\noindent This action is formulated in terms of the two-form
potential $B$ while $F=dA$ is the field-strength of a one-form
gauge potential $A$ and $*$ is the Hodge star (duality) operator.
The quadratic term is included for latter convenience.

Dimensional reduction is usually done by expanding the fields in
normal modes corresponding to the compactified extra dimensions,
and integrating out the extra dimensions. This approach is very
useful in dual models and superstrings \cite{scherk}. Here,
however, we only consider the fields in higher dimensions to be
independent of the extra dimensions.

In this case, we assume that our fields are independent of the
extra coordinate $x_3.$ From (\ref{bfm01}), on performing
dimensional reduction as described above, we get in three
dimensions

\begin{equation}
S=\int_{M_3}\left\{ B\wedge d\phi +V\wedge F
 -g^2B\wedge \,^{*}B+g^2V\wedge
\,^{*}V\right\} , \label{bfm03}
\end{equation}
where $V$ and $\phi$ are a 1-form and a 0-form fields
respectively.

We recognize that $B\wedge d\phi$ is topological in the sense that
there is no explicit dependence on the space-time metric. One has
to stress that this term may not be confused with the
two-dimensional version of the $B\wedge F$, which involves a
scalar and a one-form fields \cite{emery}. Moreover, a term that
is equivalent to the four-dimensional $B\wedge F $ term is present
in action (\ref{bfm03}) (the so-called mixed Chern-Simons term,
$V\wedge F$). On the other hand, this action displays a local
tensor gauge symmetry whose origin is to be connected to the
topological character of model and present still an invariance
under $U(1)\times U(1)$.

\section{\bf Topological Mass Generation}
\noindent Now, in order to show the topological mass generation
for the vector and tensor fields, we construct two variations from
the model (\ref{bfm03}), by introducing their propagation terms.

The model with propagation for the two-form gauge potential $B$
and with the topological term  $B\wedge d\phi$  may be represented
through the action

\begin{equation}
S=\int_{M_3}\left\{ \frac 12H\wedge \,^{*}H+\frac 12d\phi \wedge
\,^{*}d\phi +\kappa B\wedge d\phi \right\} , \label{gmt01}
\end{equation}
where the second term is a Klein-Gordon term, $\kappa $ is a mass
parameter and $H=dB$ is a three-form field-strength of $B$.

We follow here the same steps that has been used by Allen {\it{et
al.}} \cite{allen} in order to show the topological mass
generation in the context of $B\wedge F$ model. Thus, we find the
equations of motion for scalar and tensor fields, which are
respectively

\begin{equation}
d\,^{*}H=\kappa d\phi  \label{gmt03}
\end{equation}
and
\begin{equation}
d\,^{*}d\phi =-\kappa H.  \label{gmt04}
\end{equation}

Applying $d\,^{*}$ on both sides of eq. (\ref{gmt04}) and using
the
eq. (\ref{gmt03}), we obtain the equation of motion for $%
\phi $, namely

\begin{equation}
(d^{*}d^{*}+\kappa ^2)d\phi =0 . \label{gmt05}
\end{equation}
Repeating the procedure above in reverse order, we obtain the
equation of motion for $H$

\begin{equation}
(d^{*}d^{*}+\kappa ^2)H=0.  \label{gmt06}
\end{equation}

These equations can be rewritten as

\begin{equation}
(\square +\kappa ^2)\partial _\mu \phi =0  \label{gmt061}
\end{equation}
and
\begin{equation}
(\square +\kappa ^2)H=0 .  \label{gmt062}
\end{equation}
Therefore, the fluctuations of $\phi$ and $H$ are massive.
Obviously, these two possibilities can not occurs simultaneously.
Indeed, in the most interesting case, the degree of freedom of the
massless $\phi$ field is "eaten up" by the gauge field $B$ to
become massive and the $\phi$ field completely decouples from the
theory.

On the other hand, the model with propagation for two one-form
gauge fields is represented by the action

\begin{equation}
S=\int_{M_3}\left\{ \frac 12F\wedge \,^{*}F+\kappa V\wedge F+\frac
12G\wedge \,^{*}G\right\} ,  \label{gmt07}
\end{equation}
where the first term is the Maxwell one, and

\begin{equation}
G=dV.  \label{gmt08}
\end{equation}
In the equation (\ref{gmt07}) we highlight the topological term
that involves two vector fields. The equations of motion for them
are

\begin{equation}
d^{*}F=-\kappa \,dV  \label{gmt09}
\end{equation}
and
\begin{equation}
d^{*}G=\kappa \,F.  \label{gmt10}
\end{equation}

Following the former procedure we get
\begin{equation}
(\square+\kappa ^2)F=0,  \label{gmt11}
\end{equation}
and
\begin{equation}
(\square+\kappa ^2)G=0,  \label{gmt12}
\end{equation}
which shows that the fluctuations of $F$ and $G$ are massive. This
last case has already been discussed by Ghosh {\it et al.}
\cite{ghosh}.

\section{\bf A Classical Duality}
\noindent Let us take a look in the following action
\begin{equation}
\int_{M_3}\left[ B\wedge d\phi -\frac 12B\wedge \,^{*}B\right]
\label{gmt13}
\end{equation}

We would like to mention the analogy of this action with the so
called BF-Yang-Mills model \cite{fucito}. The latter formulation
take advantage of the two-form gauge field $B$ to use a first
order formalism to study pure Yang-Mills theory in four
dimensions. Furthermore, the BF-YM model (using St\"{u}ckelberg
auxiliary fields) preserves all the symmetries of the topological
$B\wedge F$ model, and so, Yang-Mills theory can be viewed as a
perturbative expansion around the pure topological theory \cite
{cattaneo}. In the present case, the Kalb-Ramond field can be also
seen as an auxiliary field, leading us to a free Maxwell action.
As a matter of fact, the non-Abelian version of (\ref{gmt13}) may
be interesting if we want to treat 3D Yang-Mills theory in a first
order formalism.

On the other hand, we can consider the action (\ref{gmt13}) as a
master equation for Klein-Gordon and Maxwell action in (2+1)
dimensions. Indeed, is easy to see that variation with respect to
$\phi $ implies that
\[
dB=0
\]
and, from the Poincar\'{e}'s lemma
\[
B=dA .
\]
Putting this result in (\ref{gmt13}), we have
\begin{equation}
S_M=-\frac 12\int_{M_3}F\wedge {\thinspace}^{*}F , \label{gmt14}
\end{equation}
which is the Maxwell action. Performing now the variation with respect to $%
B, $ we can write down
\[
B={\thinspace}^{*}d\phi.
\]
Substituting in (\ref{gmt13}) gives
\begin{equation}
S_\phi =\frac 12\int_{M_3}d\phi \wedge \,^{*}d\phi . \label{gmt15}
\end{equation}

So, from the master action (\ref{gmt13}) we have shown that $S_M$ and $%
S_\phi $ are dual to each other \cite{svend}. This duality,
specially if considered in the framework of non-Abelian
extensions, can aids greatly in unraveling interesting features of
the models with non-trivial topology. In particular, may be
interesting the study toward connection with new gauge
formulations of three-dimensional gravity \cite{jackiw}. Further
discussions about the consequences of extensions of (\ref {gmt13})
will be presented in Ref. \cite{preparation}.
\newpage
\section{\bf St\"{u}ckelberg-type Mass Generation}

\noindent We would like now to introduce a different type of mass
generation. The St\"uckelberg formulation \cite{stuck} enforces a
gauge invariance by means of an auxiliary field. The starting
point here is the gauge invariant action:

\begin{equation}
S=\int_{M_3}\left\{ \frac 12H\wedge \,^{*}H -m^2(B-d\Gamma )\wedge
\ ^{*}(\,B-d\Gamma )\right\} . \label{bfm02}
\end{equation}
This action has invariance under

\begin{equation}
B\rightarrow B+d\Omega   \label{bfm09}
\end{equation}
and
\begin{equation}
\Gamma\rightarrow \Gamma+\Omega ,  \label{bfm11}
\end{equation}

where  $\Gamma $ is the St\"{u}ckelberg one-form auxiliary field
and $\Omega $ its respective transformation parameter. The
equations of motion following from this action for $B$ and
$\Gamma$ are
\begin{equation}
d\,^{*}H-m^2\ ^{*}(\,B-d\Gamma )=0  \label{bfm16}
\end{equation}
and
\begin{equation}
d\,^{*}(\,B-d\Gamma )=0 . \label{bfm17}
\end{equation}
So the current $K=m^2\ ^{*}(\,B-d\Gamma )$ associated with $B$
field is conserved due to equation of motion of the $\Gamma$
field, such as that for Noether current.

Let us now compare two different formulation of spin-one massive
theory, namely  the actions (\ref{bfm02}) and (\ref{gmt01}). The
equation of motion from the latter action for Kalb-Ramond field is

\begin{equation}
\,^{*}d\,^{*}H=\kappa \,^{*}d\phi = J_B  . \label{bfm19}
\end{equation}
Therefore the current $J_B$  is an algebraically conserved
current, since $(\,^{*}d\,^{*})(\,^{*}d \phi) =0$ (equivalently,
$\partial^\mu J_{\mu\nu}=0$).

This interchange between topological and Noether current in two
different formulations of massive spin-one gauge theory usually
denotes a duality transformation. As we shall see now, indeed that
is the case.

Consider a global symmetry in (\ref{gmt01}) of the form $ \delta
\phi=\epsilon$ and $ \delta B=0$. The dual theory is obtained by
the procedure of gauging the  global symmetry in the model by a
one-form gauge field $A$ and constraining it to be zero by means
of a closed 2-form Lagrange multiplier $\Phi$. Thus this gauge
field is integrated and the theory is expressed in terms of the
multiplier field. This is the well-known Buscher's duality
procedure \cite{buscher}.

Having the above procedure in mind the action (\ref{gmt01}) may be
rewritten as

\begin{equation}
S=\int_{M_3}\left\{ \frac 12H\wedge \,^{*}H+\frac 12(d\phi-A)
\wedge \,^{*}(d\phi-A) +\kappa B\wedge (d\phi-A)+\kappa \Phi
\wedge A \right\} , \label{bfm18}
\end{equation}
where the 2-form field $\Phi$ is defined as $\Phi=d\Gamma$.

Deriving the action (\ref{bfm18}) respect to $A$, one gets

\begin{equation}
(d\phi - A)=-\kappa(B-\Phi) , \label{bfm20}
\end{equation}
and integrating out the field $A$ we reobtain the St\"uckelberg
action (\ref{bfm02}).

Similar results has been discussed by Harikumar and Sivakumar in
the context of four-dimensional $ B\wedge F$ model \cite{hari}.

\section{\bf Conclusions}
\noindent In summary, it was shown that a topologically massive
theory can arise from three-dimensional models containing a 2-form
(Kalb-Ramond) and a 0-form fields. Such models are obtained from
dimensional reduction of a $B\wedge F$ four-dimensional theory.
Our main analysis has focused on the role played by the
non-Chern-Simons topological term, namely, that involving a 2-form
and a 0-form fields. We showed that this term can generate mass to
the Kalb-Ramond field. Further, this term favors a classical
duality equivalence between a massless scalar field theory and a
Maxwell action.

Finally, we have shown that a 3D topologically massive theory
describing spin-one particle, with a topological term different
from the usual $B\wedge F$ term, is dually equivalent to a
St\"{u}ckelberg-type spin-one theory.

\vspace*{0.11truein}

{\bf Acknowledgments} \vspace*{0.11truein}

We wish to thank Marcony Silva Cunha for helpful discussions. This
work was supported in part by Conselho Nacional de Desenvolvimento
Cient\'{\i}fico e
Tecnol\'{o}gico-CNPq and Funda\c{c}\~{a}o Coordena\c{c}\~{a}o de Aperfei\c{c}%
oamento de Pessoal de N\'{\i}vel Superior-CAPES. D. M. M.
acknowledges the Universidade Estadual do Cear\'{a}-UECE for the
great subsidy. \vspace*{0.11truein}

%{\bf References}

\end{document}